\documentclass[runningheads]{cl2emult}

\usepackage{graphicx} 

\usepackage{amstex}   
\def\BbbR{{\Bbb R}}
\def\BbbZ{{\Bbb Z}}

\def\RPthree{{{\Bbb R \Bbb P}^3}}
\def\RPtwo{{{\Bbb R \Bbb P}^2}}

\def\AdSthree{\hbox{AdS${}_3$}}
\def\CAdSthree{\hbox{CAdS${}_3$}}
\def\Otwotwo{{\rm O}(2,2)}
\def\Othree{{\rm O}(3)}

\def\Dint{D_{\rm int}}

\def\boulvackrus{{|0_{\rm B,K}\rangle}}
\def\boulvacgeon{{|0_{\rm B,G}\rangle}}

\def\hhvackrus{{|0_{\rm K}\rangle}}
\def\hhvacgeon{{|0_{\rm G}\rangle}}

\def\Uone{{\rm U}(1)}

\def\ads{{Anti-de~Sitter}}

\def\btz{{Ba\~nados-Teitelboim-Zanelli}}

\def\hhi{{Hartle-Hawking}}

\begin{document}
\title*{Single-exterior Black Holes}
%
%
\titlerunning{\null}
\author{Jorma Louko
}
%
\authorrunning{\null}
\institute{Max-Planck-Institut f\"ur Gravitations\-physik, 
Am~M\"uhlenberg~5,\\ 
D-14476 Golm, Germany\\
$\phantom{mustanaamio}$\\
AEI 1999-002\\
gr-qc/9906031\\
$\phantom{mustanaamio}$\\
May 1999}

\maketitle   

\begin{abstract}
  We discuss quantum properties of the single-exterior, ``geon''-type
  black (and white) holes that are obtained from the Kruskal spacetime
  and the spinless \btz\ hole via a quotient construction that
  identifies the two exterior regions. For the four-dimensional geon,
  the \hhi\ type state of a massless scalar field is thermal in a
  limited sense, but there is a discrepancy between Lorentzian and
  Riemannian derivations of the geon entropy. For the
  three-dimensional geon, the state induced for a free conformal
  scalar field on the conformal boundary is similarly thermal in a
  limited sense, and the correlations in this state provide support
  for the holographic hypothesis in the context of asymptotically
  \ads\ black holes in string theory.\footnote{Lectures at
    the XXXV Karpacz Winter School on Theoretical Physics: From
    Cosmology to Quantum Gravity (Polanica, Poland, February 1999). 
To be published in the proceedings, edited by J.~Kowalski-Glikman.}
\end{abstract}

\vskip8\jot

\section{Introduction}%
\label{sec:intro}%
In quantum field theory on the Kruskal spacetime, one way to arrive at
the thermal effects is through the observation that the spacetime has
two exterior regions separated by a bifurcate Killing horizon.  A~free
scalar field on the Kruskal spacetime has a vacuum state, known as the
\hhi\ vacuum~\cite{hh-vacuum,israel-vacuum}, that is invariant under
all the continuous isometries of the
spacetime~\cite{kay-wald,wald-qft}.  This state is pure, but the
expectation values of operators with support in {\em one\/} exterior
region are thermal in the Hawking
temperature~\cite{hh-vacuum,israel-vacuum,kay-wald,wald-qft,unruh-magnum}.
Similar observations hold for field theory on the nonextremal
(2+1)-dimensional \btz\ (BTZ) black hole, both with and without
spin~\cite{carlip-rev}, and also for conformal field theory on the
conformal boundary of the BTZ
hole~\cite{MAST,horo-marolf-strc,louko-marolf-btz}.

In all these cases one has a vacuum state that knows about the
global geometry of the spacetime, in particular about the fact that the
spacetime has two exterior regions.  
Suppose now that we modify the spacetime in
some  `reasonable' fashion so that one exterior region remains as it is,
and  all the modification takes place behind the Killing horizons of this
exterior region. Suppose further that the modified spacetime admits a
vacuum state that is, in some reasonable sense, a 
\hhi\ type vacuum. Can we then, by
probing the new vacuum in the unmodified exterior region, 
discover
that something has happened to the spacetime behind the horizons? 
In particular, as the new spacetime 
is still a black (and white) hole, does the new vacuum exhibit
thermality, and if so, at what temperature? 
In the
(2+1)-dimensional case, the analogous questions can also 
be raised for
conformal field theory on the conformal boundary. 

These lectures address the above questions for a particular
modification of the Kruskal manifold and the spinless BTZ hole: we
modify the spacetimes by a quotient construction that identifies the
two exterior regions with each other. 
For Kruskal, the resulting spacetime is referred to as the $\RPthree$ geon
\cite{Nico,topocen}, and for BTZ, as the $\RPtwo$
geon~\cite{louko-marolf-btz}. These spacetimes are black (and white)
holes, and their only singularities are those inherited from the
singularities of the two-exterior holes. 

On the $\RPthree$ geon, a free scalar field has a vacuum induced from
the \hhi\ vacuum on Kruskal. The vacuum is not fully thermal for
static exterior observers, but it appears thermal when probed with
operators that do not see certain types of correlations, such as in
particular operators with support at asymptotically late times, and
the apparent temperature is then the usual Hawking temperature.
However, a~naive application of Euclidean-signature path-integral
methods via saddle-point methods yields for the geon only half of the
Bekenstein-Hawking entropy of the Schwarzschild hole with the same
mass.

The situation on the 
conformal boundary of the $\RPtwo$ geon is
analogous. The quotient construction from the conformal
boundary of \ads\ space induces on the boundary of the geon a 
\hhi\ type vacuum that is not fully thermal, but it appears thermal
when probed with operators that do not see certain types of
correlations, and the apparent temperature is then the usual Hawking
temperature of the BTZ hole. 
The properties of the boundary vacuum turn out 
to reflect in a surprisingly
close fashion the geometry of the geon spacetime. 
This can be interpreted as support for the holographic hypothesis
\cite{thooft-holo,susskind-holo}, according to which physics in the
bulk of a spacetime should be retrievable from physics on the boundary
of the spacetime. It further suggests that single-exterior black holes
can serve as a test bed for the versions of the
holographic hypothesis that arise in string theory 
for asymptotically \ads\ 
spacetimes via the Maldacena duality 
conjectures~\cite{MAST,MAL,malda-more,malda-stillmore}. 

The material is based on joint work 
\cite{louko-marolf-btz,louko-marolf-geon} with Don Marolf, whom I would
like to thank for a truly delightful collaboration. I~would also like to
thank the organizers of the Polanica Winter School for the opportunity to
present the work in a most pleasant and inspiring atmosphere.

\section{Kruskal Manifold and the $\RPthree$ Geon}%
\begin{figure}
\begin{center}
\includegraphics[width=.7\textwidth]{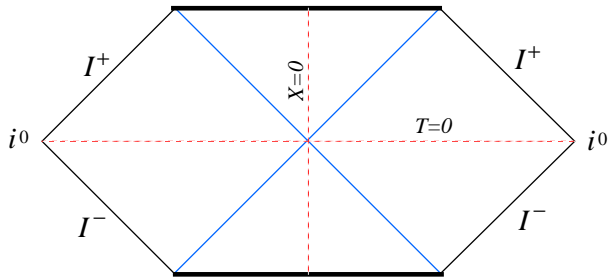}
\end{center}
\caption[]{Conformal diagram of the Kruskal spacetime.
Each point represents a suppressed $S^2$ orbit
of the $\Othree$ isometry group
}
\label{fig:Kruskal}
\end{figure}

\label{sec:kruskal-and-geon}%
Recall that the metric on the Kruskal manifold ${\cal M}^L$ reads 
\begin{equation}
ds^2 =
{32M^3 \over r} \exp\left(-\frac{r}{2M}\right)
\left( -dT^2 + dX^2 \right)
+ r^2 d\Omega^2
\; , 
\label{Kruskal-metric-lor}
\end{equation}
where $d\Omega^2 = d\theta^2 + \sin^2\theta \, d\varphi^2$ 
is the metric
on the unit two-sphere, 
$M>0$, 
$X^2 - T^2 > -1$, 
and $r$ is determined as a function of $T$ and $X$ by 
\begin{equation}
\left(
{r \over 2M} -1
\right)
\exp \! \left(\frac{r}{2M}\right)
=
X^2 -T^2
\; .
\label{r-on-Kruskal}
\end{equation}
The coordinates are global, 
apart from the elementary singularities of the spherical coordinates. 
${\cal M}^L$~is manifestly spherically symmetric, 
and it has in addition the 
Killing vector
\begin{equation}
V^L :=
\frac{1}{4M}
\left(
X \partial_T + T \partial_X
\right)
\; ,
\label{KillingV-lor}
\end{equation}
which is timelike for $|X| > |T|$ and spacelike for $|X| < |T|$. 
A~conformal diagram of ${\cal M}^L$, with the two-spheres suppressed,
is shown in Fig.~\ref{fig:Kruskal}.

In each of the four quadrants of ${\cal M}^L$ one
can introduce Schwarzschild coordinates
$(t,r,\theta,\varphi)$ that are adapted to 
the isometry generated by~$V^L$. In the ``right-hand-side'' exterior
region, $X > |T|$, the coordinate transformation reads 
\begin{eqnarray}
T
&=&
\left(\frac{r}{2M}-1\right)^{1/2}
\exp \! \left(\frac{r}{4M}\right)
\sinh\left(\frac{t}{4M}\right)
\; ,
\nonumber
\\
X
&=&
\left(\frac{r}{2M}-1\right)^{1/2}
\exp \! \left(\frac{r}{4M}\right)
\cosh\left(\frac{t}{4M}\right)
\; ,
\label{Sch-lor-coords}
\end{eqnarray}
with $r>2M$ and $-\infty<t<\infty$. 
The exterior
metric takes then the Schwarzschild form 
\begin{equation}
ds^2 =
- \left(1 - \frac{r}{2M}\right) dt^2
+
{dr^2 \over
{\displaystyle{\left(1 - \frac{r}{2M}\right)}}}
+ r^2 d\Omega^2
\; , 
\label{Sch-metric-lor}
\end{equation}
and $V^L = \partial_t$. 

\begin{figure}
\begin{center}
\includegraphics[width=.5\textwidth]{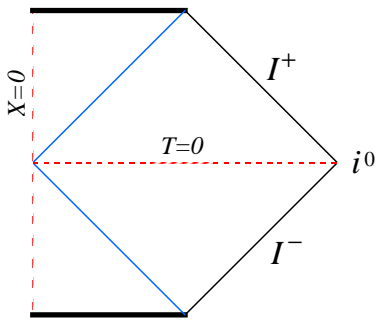}
\end{center}
\caption[]{Conformal diagram of the $\RPthree$ geon ${\cal
M}^L/J^L$. Each point represents a suppressed orbit
of the $\Othree$ isometry group.
The region $X>0$ is isometric to the region
$X>0$ of~${\cal M}^L$,
shown in Fig.~\ref{fig:Kruskal}, and 
the $\Othree$ isometry orbits
in this region are two-spheres.
At $X=0$, the $\Othree$ orbits have topology~$\RPtwo$
}
\label{fig:RPthree}
\end{figure}

Consider now on ${\cal M}^L$ the isometry 
\begin{equation}
J^L:
(T,X,\theta,\varphi)
\mapsto
(T,-X,\pi-\theta,\varphi+\pi)
\; . 
\label{JL-in-Kruskalcoords}
\end{equation}
$J^L$ is clearly involutive, it acts properly discontinuously, 
it preserves the time orientation and 
spatial orientation, and it commutes with the spherical symmetry
of~${\cal M}^L$. The quotient space ${\cal
M}^L/J^L$ 
is therefore a spherically symmetric, space and time
orientable manifold. 
A~conformal diagram of ${\cal
M}^L/J^L$ is shown in Fig.~\ref{fig:RPthree}.
${\cal M}^L/J^L$
is an inextendible black (and white) hole spacetime, 
and its only singularities
are those inherited from the singularities of~${\cal M}^L$. 
It has only one exterior region, and its spatial topology is 
$\RPthree\setminus${}$\{$point at infinity$\}$. 
We refer to ${\cal M}^L/J^L$ as the $\RPthree$
geon~\cite{Nico,topocen}. 

The exterior region of ${\cal M}^L/J^L$ is clearly isometric to an 
exterior region of~${\cal M}^L$. In terms of the coordinates
shown in Fig.~\ref{fig:RPthree}, the exterior region is at 
$X > |T|$, and one can introduce 
in the exterior region 
standard Schwarzschild
coordinates by~(\ref{Sch-lor-coords}). 
As the Killing
vector $V^L$ on ${\cal M}^L$ changes its sign under~$J^L$, the
timelike 
Killing vector $\partial_t$ 
on the exterior of ${\cal M}^L/J^L$ 
can however not be continued into a
globally-defined Killing vector on ${\cal M}^L/J^L$. This means
that not all the constant $t$ hypersurfaces in the exterior region of
${\cal M}^L/J^L$ are equal: among them, there is only one 
(in Fig.~\ref{fig:RPthree}, the one at $T=0$) 
that can be extended into a
smoothly-embedded Cauchy hypersurface for ${\cal M}^L/J^L$. 

The quotient construction from ${\cal M}^L$ to ${\cal M}^L/J^L$ can be
analytically continued to the Riemannian 
(i.e., positive definite) sections 
via the formalism of (anti-)holomorphic
involutions~\cite{gibb-holo,chamb-gibb}. The 
Riemannian section of the Kruskal hole, denoted by ${\cal M}^R$, is
obtained from 
(\ref{Kruskal-metric-lor}) and (\ref{r-on-Kruskal}) by setting 
$T = -i {\tilde T}$ and letting ${\tilde T}$ and $X$ take all real
values~\cite{GH1}. 
The analytic continuation of~$J^L$, denoted by~$J^R$, acts on ${\cal
  M}^R$ by 
\begin{equation}
J^R:
({\tilde T},X,\theta,\varphi)
\mapsto
({\tilde T},-X,\pi-\theta,\varphi+\pi)
\; ,
\label{JR-in-Kruskalcoords}
\end{equation}
and the Riemannian section of the $\RPthree$ geon is
${\cal M}^R/J^R$. 

\begin{figure}
\begin{center}
\includegraphics[width=.9\textwidth]{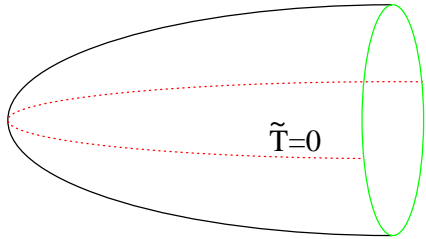}
\end{center}
\caption[]{A~``sock'' 
  representation of the Riemannian section ${\cal M}^R$ of the
  complexified Kruskal manifold. 
  The $S^2$ orbits of the $\Othree$ isometry group are
  suppressed, and the remaining two dimensions $({\tilde{T}},X)$ are
  shown as an isometric embedding into Euclidean~$\BbbR^3$.  The
  isometry generated by $\partial_{{\tilde{t}}}$ rotates the
  two shown dimensions
}
\label{fig:riemkruskal}
\end{figure}

\begin{figure}
\begin{center}
\includegraphics[width=.9\textwidth]{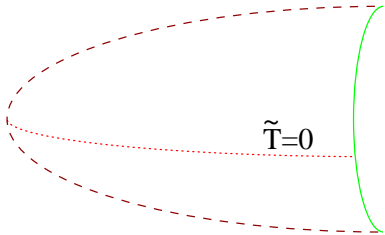}
\end{center}
\caption[]{A~representation of the Riemannian 
  section ${\cal M}^R/J^R$ of the complexified $\RPthree$ geon as the
  ``front half'' of the the ${\cal M}^R$ sock.  The orbits of the
  $\Othree$ isometry group are suppressed, as in
  Fig.~\ref{fig:riemkruskal}.  The generic orbits have topology~$S^2$,
  but those at the ``boundary'' of the diagram ({\it dashed line\/})
  have topology~$\RPtwo$ 
}
\label{fig:riemrpthree}
\end{figure}

On ${\cal M}^R$ we can introduce the 
Riemannian Schwarzschild coordinates 
$({\tilde{t}},r,\theta,\varphi)$, obtained from the
Lorentzian Schwarzschild coordinates for $r>2M$ by 
$t=-i {\tilde{t}}$. These Riemannian Schwarzschild coordinates are
global, with the exception of a coordinate singularity at the
Riemannian horizon $r=2M$, provided they are understood with the 
identification 
$({\tilde{t}},r,\theta,\varphi) \sim 
({\tilde{t}} + 8\pi M , r,\theta,\varphi)$~\cite{GH1}. 
On ${\cal M}^R/J^R$, 
the Riemannian Schwarzschild 
coordinates need to be understood with the additional 
identification $({\tilde{t}},r,\theta,\varphi)
\sim 
({\tilde{t}} + 4\pi M, r,\pi-\theta,\varphi+\pi)$, which arises from the
action (\ref{JR-in-Kruskalcoords}) of $J^R$ on~${\cal M}^R$. 
The Killing vector 
$\partial_{\tilde{t}}$ is global on~${\cal M}^R$, 
and it generates an
$\Uone$ isometry group with a fixed point at the Riemannian
horizon. On ${\cal M}^R/J^R$, on the other hand,
$\partial_{\tilde{t}}$ is global only as a line field but not as a
vector field, and the analogous 
$\Uone$ isometry does not exist. 
Embedding diagrams of ${\cal M}^R$ and ${\cal
  M}^R/J^R$, with the orbits of the spherical symmetry suppressed, 
are shown in Figs.\ \ref{fig:riemkruskal} and~\ref{fig:riemrpthree}. 

\section{Vacua on
Kruskal and on the $\RPthree$ Geon}
\label{sec:hhivacs}

We now consider a free scalar field on the Kruskal manifold and on the
$\RPthree$ geon. For concreteness, we take here the field to be
massless. The situation with a massive field is qualitatively 
similar~\cite{louko-marolf-geon}.

Recall that the \hhi\ vacuum $\hhvackrus$ of a massless scalar
field on the Kruskal manifold ${\cal M}^L$ can be characterized by its
positive frequency properties along the affine parameters of the
horizon generators \cite{hh-vacuum,israel-vacuum,unruh-magnum}, by the
complex analytic properties of the Feynman propagator upon analytic
continuation to ${\cal M}^R$ \cite{hh-vacuum}, or by the invariance
under the continuous isometries of ${\cal M}^L$~\cite{kay-wald,wald-qft}. 
$\hhvackrus$~is regular everywhere on ${\cal
  M}^L$, but it is not annihilated by the annihilation operators
associated with the future timelike Killing vectors in the exterior 
regions: a static observer in an exterior region sees $\hhvackrus$ as
an excited state. We have the expansion 
\begin{equation}
\hhvackrus = \sum_{i \cdots k} f_{i \cdots k}
\left( a_i^R \right)^\dagger 
\left( a_i^L \right)^\dagger 
\cdots 
\left( a_k^R \right)^\dagger 
\left( a_k^L \right)^\dagger 
\boulvackrus
\; ,
\label{hhvackrus-expansion}
\end{equation}
where the Boulware vacuum $\boulvackrus$ is the vacuum with respect to the 
timelike Killing vectors in the exterior regions, 
$\left( a_i^R \right)^\dagger$ are the creation operators with respect to 
this Killing vector in the right-hand-side exterior region, and 
$\left( a_i^L \right)^\dagger$ are the creation operators with respect to 
this Killing vector in the left-hand-side exterior region. 

$\hhvackrus$~thus contains Boulware
excitations in correlated pairs, such that one member of the pair has
support in the right-hand-side exterior and the other member in 
the left-hand-side exterior. An operator with support in (say) the
right-hand-side exterior does not couple to the left-hand-side
excitations, and the expectation values of such operators in
$\hhvackrus$ thus look like expectation values in a mixed state. 
{}From the detailed form of the expansion
coefficients $f_{i \cdots k}$ (which we do not write out
here) it is seen that this mixed state 
is thermal, and it 
has at infinity the Hawking temperature
$T = {(8\pi M)}^{-1}$. 

Now, through the quotient construction from ${\cal M}^L$ to ${\cal
  M}^L/J^L$, 
$\hhvackrus$ induces on ${\cal
  M}^L/J^L$ a \hhi\ type vacuum, which we denote by~$\hhvacgeon$. 
Again, $\hhvacgeon$ can be characterized by its 
positive frequency properties along the affine parameters of the
horizon generators, or by the
complex analytic properties of the Feynman 
propagator~\cite{louko-marolf-geon}. $\hhvacgeon$~has the expansion 
\begin{equation}
\hhvacgeon = \sum_{i \cdots k} {\tilde f}_{i \cdots k}
\left( {\tilde a}_i^{(1)} \right)^\dagger 
\left( {\tilde a}_i^{(2)} \right)^\dagger 
\cdots 
\left( {\tilde a}_k^{(1)} \right)^\dagger 
\left( {\tilde a}_k^{(2)} \right)^\dagger 
\boulvacgeon
\; ,
\label{hhvacgeon-expansion}
\end{equation}
where $\boulvacgeon$ is the Boulware vacuum in the single exterior
region and $\left( {\tilde a}_i^{(\alpha)} \right)^\dagger$ are the 
creation operators of Boulware particles in the exterior region. The
indices $i$ and $\alpha$ now label a complete set of positive
frequency Boulware modes in the single exterior region.

We see from (\ref{hhvacgeon-expansion}) that $\hhvacgeon$
contains Boulware excitations in correlated pairs, but the crucial
point is that both members of each pair have support in the single
exterior region. Consequently, the expectation values of arbitrary
operators in the exterior region are not thermal.  However, for
operators that do not contain couplings between modes with $\alpha=1$
and $\alpha=2$, the expectation values turn out to be thermal, with
the Hawking temperature $T = {(8\pi M)}^{-1}$. One class of operators
for which this is the case are operators with, roughly speaking, 
support at
asymptotically late (or early) times: the reason is that an excitation 
with support at asymptotically 
late exterior times is correlated with one with
support at asymptotically early exterior times. Note that ``early''
and ``late'' here mean compared with the distinguished 
exterior spacelike hypersurface
mentioned in
Sect.~\ref{sec:kruskal-and-geon} 
(in Fig.~\ref{fig:RPthree}, the one at $T=0$). 

Thus, for a late-time observer in the exterior region of ${\cal
  M}^L/J^L$, the state $\hhvacgeon$ is indistinguishable from the
state $\hhvackrus$ on~${\cal M}^L$. This conclusion can also be 
reached by 
analyzing the response of a monopole particle detector, or from an
emission-absorption analysis analogous to that performed for
$\hhvackrus$ in~\cite{hh-vacuum}, provided certain technical
assumptions about the falloff of the two-point functions in 
$\hhvacgeon$ hold~\cite{louko-marolf-geon}.

\section{Entropy of the $\RPthree$ Geon?}%
\label{sec:geon-entropy}%
As explained above, for a late-time exterior observer in ${\cal
  M}^L/J^L$ the state $\hhvacgeon$ is indistinguishable from the
state $\hhvackrus$ on~${\cal M}^L$. The late-time observer can
therefore promote the 
classical first law of black hole mechanics \cite{BarCarHaw}
into a first law of black hole thermodynamics exactly as for the
Kruskal black hole~\cite{hawkingCMP,bekenstein1,bekenstein2}. The
observer thus finds for the thermodynamic late time entropy of
the geon the usual Kruskal value~$4 \pi M^2$, which is one
quarter of the area of the geon black hole horizon at late times.  
If one views the geon as a dynamical black-hole spacetime, with the
asymptotic far-future horizon area~$16 \pi M^2$, this is the result
one might have expected on physical grounds.

On the other hand, the area-entropy relation for the geon is made
subtle by the fact that the horizon area is not constant along
the horizon. Away from the intersection of the past and future
horizons, the horizon duly has topology $S^2$ and area $16 \pi M^2$,
just as in Kruskal. The critical surface at the intersection of the
past and future horizons, however, has topology $\RPtwo$ and area~$8
\pi M^2$. As it is precisely this critical surface that belongs to
both the Lorentzian and Riemannian sections of the complexified
manifold, and constitutes the horizon of the Riemannian section, one
may expect that methods utilizing the Riemannian section of the
complexified manifold 
\cite{GH1,hawkingCC}
produce
for the geon entropy the value~$2 \pi M^2$, 
which is one quarter of the
critical surface area, and only half of the Kruskal entropy. This
indeed is the case, provided the surface terms in the
Riemannian geon action are handled in a way suggested by the quotient
construction from ${\cal M}^R$ to 
${\cal M}^R/J^R$~\cite{louko-marolf-geon}. 

There are several possible physical interpretations for this 
disagreement between the Lorentzian and Riemannian results for the
entropy. At one extreme, it could be that the 
path-integral framework
is simply 
inapplicable to the geon, for reasons having to do with 
the absence of certain globally-defined 
symmetries. 
For instance, despite the fact that the exterior region of ${\cal
  M}^L/J^L$ is static, the restriction of $\hhvacgeon$ to this region
is not.  Also, the asymptotic region of ${\cal M}^R/J^R$ does not have
a globally-defined Killing field, 
and the homotopy group of any neighborhood of
infinity in ${\cal M}^R/J^R$ is $\BbbZ_2$ as opposed to the trivial
group.  It may well be that such an asymptotic structure does not
satisfy the boundary conditions that should be imposed in the path
integral for the quantum gravitational partition function.

At another extreme, 
it could be that the path-integral framework is applicable
to the geon, and our way of applying it is correct, but the resulting
entropy is physically distinct from the subjective thermodynamic
entropy associated with the late-time exterior observer.  If this is
the case, the physical interpretation of the path-integral entropy
might be in the quantum statistics in the whole
exterior region, and one might anticipate
this entropy 
to arise from tracing over degrees of freedom that
are in some sense unobservable. It would thus be
interesting to see 
see whether any state-counting calculation for the geon entropy 
would produce agreement with the path-integral result.

\section{$\AdSthree$, the Spinless Nonextremal BTZ Hole, 
\\
and the $\RPtwo$ Geon}%
\label{sec:btz-and-geon}%
We now turn to 2+1 spacetime dimensions. In this section we review how 
the spinless nonextremal BTZ hole and the $\RPtwo$ geon arise as
quotient spaces of the three-dimensional \ads\ space, and how this
quotient construction can be extended to the conformal boundaries. 

\subsection{$\AdSthree$, its Covering Space, 
and the Conformal Boundary}%
\label{subsec:adsthree}%
Recall that the three-dimensional \ads\ space 
($\AdSthree$) can be
defined as the hyperboloid 
\begin{equation}
- 1 = - {(T^1)}^2 - {(T^2)}^2 
+ {(X^1)}^2 + {(X^2)}^2 
\label{embedding-surface}
\end{equation}
in $\BbbR^{2,2}$ with the metric 
\begin{equation}
ds^2 = 
- {(dT^1)}^2 - {(dT^2)}^2 
+ {(dX^1)}^2 + {(dX^2)}^2 
\; .
\label{embedding-metric}
\end{equation}
We have here normalized the Gaussian curvature of $\AdSthree$ 
to~$-1$. 
This embedding representation makes transparent the fact that 
$\AdSthree$ is a maximally symmetric space with the isometry group 
$\Otwotwo$. 

For understanding the structure of the infinity, we 
introduce the coordinates 
$(t,\rho,\theta)$ by \cite{ABBHP}
\begin{eqnarray}
&& T^1 = 
\frac{1+\rho^2}{1-\rho^2} \cos t
\; ,
\ \ \ 
T^2 = 
\frac{1+\rho^2}{1-\rho^2} \sin t
\; ,
\nonumber
\\
&&X^1 =
\frac{2\rho}{1-\rho^2} \cos \theta
\; ,
\ \ \ 
X^2 = 
\frac{2\rho}{1-\rho^2} \sin \theta
\; .
\label{sausage-coords}
\end{eqnarray}
With $0\le \rho<1$ and 
the identifications 
$(t,\rho,\theta) \sim 
(t,\rho,\theta+2\pi)
\sim 
(t+2\pi,\rho,\theta)$, 
these coordinates can be understood as global on $\AdSthree$, apart
from the elementary coordinate singularity at $\rho=0$. The metric
reads
\begin{equation}
ds^2 = \frac{4}{{(1-\rho^2)}^2}
\left[
- \frac{1}{4}
{(1+\rho^2)}^2 
dt^2 
+ d\rho^2 + \rho^2 d\theta^2
\right]
\; .
\label{sausage-metric}
\end{equation}
Dropping now from (\ref{sausage-metric}) the conformal factor
$4{(1-\rho^2)}^{-2}$ yields a spacetime that can be regularly extended
to $\rho=1$, and the timelike hypersurface $\rho=1$ in this conformal
spacetime is by definition the conformal boundary of $\AdSthree$. It
is a timelike two-torus 
coordinatized by $(t,\theta)$ with the identifications
$(t,\theta) \sim 
(t,\theta+2\pi)
\sim 
(t+2\pi,\theta)$, 
and it has the flat metric 
\begin{equation}
ds^2 = -dt^2 + d\theta^2
\; .
\label{boundary-metric}
\end{equation}

The conformal boundary 
construction generalizes in an obvious way to the universal
covering space of $\AdSthree$, which we denote by~$\CAdSthree$. 
The only difference is that the coordinate $t$ 
is not periodically identified. The conformal
boundary of~$\CAdSthree$, which we denote by~$B_C$, 
is thus a timelike cylinder with the metric 
(\ref{boundary-metric}) and the identification 
$(t,\theta) \sim (t,\theta+2\pi)$. 

\subsection{The Spinless Nonextremal BTZ Hole}%
\begin{figure}
\begin{center}
\includegraphics[width=.4\textwidth]{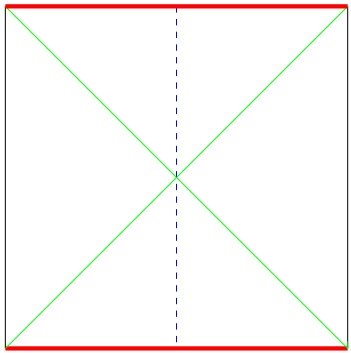}
\end{center}
\caption[]{A~conformal diagram of the BTZ
  hole. Each point in the diagram represents a suppressed~$S^1$. The
  involution ${\tilde J}_{\rm int}$ introduced in Subsect.\ 
  \ref{subsec:rptwogeon} consists of a left-right reflection about the
  dashed vertical line, followed by a rotation by $\pi$ on the
  suppressed~$S^1$
}
\label{fig:btz}
\end{figure}

\label{subsec:btz}%
Let $\xi_{\rm int}$ be on $\CAdSthree$ the Killing vector
induced by the boost-like Killing vector 
$\xi_{\rm emb} := 
- T^1 \partial_{X^1} 
- X^1 \partial_{T^1}$
of~$\BbbR^{2,2}$, and let $\Dint$ denote the largest subset of
$\CAdSthree$ that contains the 
hypersurface $t=0$ and in which $\xi_{\rm int}$ is
spacelike. Given a prescribed positive parameter~$a$, 
the isometry 
$\exp(a\xi_{\rm int})$ generates
a discrete isometry group 
$\Gamma_{\rm int}\simeq \BbbZ$ of $\Dint$. The spinless nonextremal
BTZ hole is by definition the quotient space $\Dint/\Gamma_{\rm
  int}$~\cite{carlip-rev,BTZ}. 
A~conformal diagram, with the $S^1$ factor arising from the
identification suppressed, is shown in Fig.~\ref{fig:btz}. 
The horizon circumference is~$a$, and the ADM mass
is $M=a^2/(32\pi^2 G_3)$, where $G_3$ is the (2+1)-dimensional
Newton's constant.  
For further 
discussion, including expressions for the metric in coordinates
adapted to the isometries, we refer to~\cite{carlip-rev,BTZ}. 

As seen in Fig.~\ref{fig:btz}, the BTZ hole has two exterior regions, 
and the infinities are asymptotically
\ads. The point of interest for us is that each of the infinities
has a conformal boundary that is induced from $B_C$
by the quotient construction. Technically, one
observes that $\xi_{\rm int}$ induces on $B_C$ 
the conformal Killing vector $\xi := 
\cos t \sin\theta \, \partial_\theta
+ \sin t \cos\theta \, \partial_t$, and that 
$\Dint$ reaches $B_C$ 
in the two diamonds
\begin{eqnarray}
&&
D_R := \left\{ (t,\theta) 
\mid 
\hbox{$0<\theta<\pi$, $|t| < \pi/2 - |\theta - \pi/2|$}
\right\}
\; ,
\nonumber
\\
&&
D_L := \left\{ (t, \theta) 
\mid 
\hbox{$-\pi<\theta<0$, $|t| < \pi/2 - |\theta + \pi/2|$}
\right\}
\; . 
\label{DRLdefs}
\end{eqnarray}
The two conformal 
boundaries of the BTZ hole are then the quotient spaces 
$D_R/\Gamma_R$ and~$D_L/\Gamma_L$, where $\Gamma_R$ and $\Gamma_L$ are
the restrictions to respectively $D_R$ and $D_L$ of the conformal
isometry group of $B_C$ generated by
$\exp(a\xi)$~\cite{MAST,horo-marolf-strc}. 
To make this explicit, we 
cover $D_R$ by the coordinates 
\begin{eqnarray}
\alpha 
&=&
- \ln \tan\left[(\theta-t)/2\right]
\; ,
\nonumber
\\
\beta
&=&
\ln \tan\left[(\theta+t)/2\right]
\; ,
\label{alphabeta-R}
\end{eqnarray}
in which the metric 
induced from (\ref{boundary-metric}) is conformal to 
\begin{equation}
ds^2 = - {\left(\frac{2\pi}{a}\right)}^2 
\, 
d\alpha \, d\beta
\; ,
\label{DR-metric1-conf}
\end{equation}
and 
$\xi = -\partial_\alpha + \partial_\beta$. 
The quotient space $D_R/\Gamma_R$, with the metric induced
from~(\ref{DR-metric1-conf}), is thus isometric to $B_C$ with the
metric~(\ref{boundary-metric}). In particular, it has topology
$\BbbR\times S^1$. 
It can be shown 
\cite{MAST,horo-marolf-strc,louko-marolf-btz}
that the
conventionally-normalized Killing
vector of the BTZ hole that is timelike in the exterior regions
induces on $D_R/\Gamma_R$ the timelike Killing vector 
$\eta = \partial_\alpha + \partial_\beta$. 
Analogous observations apply to~$D_L/\Gamma_L$. 

\subsection{The $\RPtwo$ Geon}%
\begin{figure}
\begin{center}
\includegraphics[width=.2\textwidth]{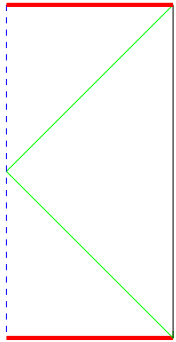}
\end{center}
\caption[]{A~conformal diagram of the $\RPtwo$ geon. 
The region not on the dashed
line is identical to that in the diagram of Fig.~\ref{fig:btz}, 
each point 
representing a suppressed $S^1$ in the spacetime. On the dashed
line, each point in the diagram represents again an $S^1$ in the
spacetime, but with only half of the circumference 
of the $S^1$'s in the
diagram of Fig.~\ref{fig:btz}
}
\label{fig:rp2geon}
\end{figure}

\label{subsec:rptwogeon}%
The $\RPtwo$ geon is obtained from the spinless BTZ hole in close
analogy with the quotient construction used with the $\RPthree$ geon
in Sect.~\ref{sec:kruskal-and-geon}. We denote the relevant
involutive isometry of the BTZ hole by ${\tilde J}_{\rm int}$: in the
conformal diagram of Fig.~\ref{fig:btz}, ${\tilde J}_{\rm int}$
consists of a left-right reflection about the dashed vertical line,
followed by a rotation by $\pi$ on the suppressed~$S^1$.  A~conformal
diagram of the quotient space, the $\RPtwo$ geon, is shown in
Fig.~\ref{fig:rp2geon}.  It is clear that the $\RPtwo$ geon is a black
(and white) hole spacetime with a single exterior region that is
isometric to one exterior region of the BTZ hole. It is time
orientable but not space orientable, and the spatial topology is
$\RPtwo\setminus${}$\{$point at infinity$\}$. The local and global
isometries closely parallel those of the $\RPthree$
geon~\cite{louko-marolf-btz}.

The map ${\tilde J}_{\rm int}$ can clearly be extended to the
conformal boundary of the BTZ hole, where it defines an involution
${\tilde J}$ that interchanges the two boundary components. Quotienting 
the conformal boundary of the BTZ hole by this involution gives the 
conformal boundary of the $\RPtwo$ geon, which is thus isomorphic to
one boundary component of the  BTZ hole. 
Note that although the $\RPtwo$ 
geon is not
space orientable, its conformal boundary 
$\BbbR \times S^1$ is.

\section{Vacua on the Conformal Boundaries}%
\label{sec:c-vacua}%
We now turn to a free conformal scalar field on the boundaries of
$\CAdSthree$, the BTZ hole, and the $\RPtwo$ geon.  

Let $|0\rangle$
denote on $B_C$ 
the vacuum state with respect to the
timelike Killing vector~$\partial_t$. We wish to know what kind of
states $|0\rangle$ induces on the conformal boundaries of the BTZ hole
and the $\RPtwo$ geon. For concreteness, we focus the presentation
on the non-zero modes of the field. The subtleties with the
zero-modes are discussed in~\cite{louko-marolf-btz}. 

Consider the boundary 
of the BTZ hole. As noted above, 
the timelike Killing vectors on the two
components do not lift to the timelike
Killing vector $\partial_t$ on~$B_C$: the future timelike Killing
vector on $D_R/\Gamma_R$ lifts to
$[a/(2\pi)]\eta$, and an analogous statement holds 
for~$D_L/\Gamma_L$. 
To interpret 
the state induced by $|0\rangle$ on
the BTZ hole boundary
in terms of the BTZ particle modes, 
we must first first write the state induced by
$|0\rangle$ on $D_R\cup D_L$ in terms of continuum-normalized particle
states that are positive frequency with respect to $\eta$ on $D_R$ and 
with respect to the analogous Killing vector on~$D_L$, and then
restrict to appropriately periodic field modes in order to accommodate 
the identification by~$\exp(a\xi)$. This calculation is quite similar
to expressing the Minkowski vacuum in terms of 
Rindler particle modes~\cite{wald-qft,birrell-davies,takagi-magnum}. 
Denoting 
the state induced from $|0\rangle$ by 
$|\hbox{BTZ}\rangle$, 
we have the expansion 
\begin{equation}
|\hbox{BTZ}\rangle
= \sum_{i \cdots k} f_{i \cdots k}
\left( a_i^R \right)^\dagger 
\left( a_i^L \right)^\dagger 
\cdots 
\left( a_k^R \right)^\dagger 
\left( a_k^L \right)^\dagger 
|0\rangle_R \, |0\rangle_L
\; ,
\label{btzvac-exp}
\end{equation}
where $|0\rangle_R$ and $|0\rangle_L$ are respectively the vacua on
the two boundary components with respect to their timelike Killing
vectors, $\left( a_i^R \right)^\dagger$ 
are the creation operators with
respect to this Killing vector on~$D_R/\Gamma_R$, 
and $\left( a_i^L \right)^\dagger$ are the creation operators
with respect to this Killing vector on~$D_L/\Gamma_L$. 
The analogy to the expansion (\ref{hhvackrus-expansion})
of the \hhi\ vacuum on Kruskal is clear: the excitations come in
correlated pairs, the two members of each pair now living on different
boundary components. Restriction to one boundary component yields a
thermal state, and when the normalization of the boundary timelike
Killing vector is matched to that in the bulk of the spacetime,
the temperature 
turns out to be the Hawking temperature of the BTZ
hole, $a/4\pi^2$. This is the result first found in~\cite{MAST}. 

The boundary of the $\RPtwo$ geon has a single connected
component. 
Denoting the state induced from $|0\rangle$ 
by $|\hbox{$\RPtwo$}\rangle$, 
we have the expansion 
\begin{equation}
|\hbox{$\RPtwo$}\rangle
= \sum_{i \cdots k} {\tilde f}_{i \cdots k}
\left( {\tilde a}_i^{(+)} \right)^\dagger 
\left( {\tilde a}_i^{(-)} \right)^\dagger 
\cdots 
\left( {\tilde a}_k^{(+)} \right)^\dagger 
\left( {\tilde a}_k^{(-)} \right)^\dagger 
|0\rangle_R
\; ,
\end{equation}
where $|0\rangle_R$ now denotes the geon boundary 
vacuum with respect 
to the timelike Killing
vector, $\left( {\tilde a}_i^{(\alpha)} \right)^\dagger$ are 
the creation
operators with respect to this Killing vector, and the indices $i$ and
$\alpha$ label the modes. 
The modes with $\alpha=+$ are right-movers 
and the modes with $\alpha=-$
are left-movers. 
The analogy to the expansion 
(\ref{hhvacgeon-expansion}) of the \hhi\ type vacuum on the $\RPthree$ 
geon is clear. For
operators that do not contain couplings between modes with $\alpha=+$
and $\alpha=-$, the expectation values turn out to be thermal, with
the BTZ Hawking temperature~$a/4\pi^2$. 

As shown in Table~\ref{table:comparison}, 
several properties of the state 
$|\hbox{$\RPtwo$}\rangle$
reflect
properties of the $\RPtwo$ geon spacetime geometry.
First, $|\hbox{$\RPtwo$}\rangle$ is a {\em pure\/} state on
the boundary cylinder $\BbbR\times S^1$: 
this follows by construction since 
(unlike with the BTZ hole) the single cylinder constitutes
the whole conformal boundary. 
Second, $|\hbox{$\RPtwo$}\rangle$ is an excited state with respect to
the boundary 
timelike Killing field. 
This can be understood to reflect the fact that 
the spacetime attached to the boundary is not~$\CAdSthree$. 
Third, it
can be shown that $|\hbox{$\RPtwo$}\rangle$
is not invariant under
translations generated by the timelike Killing vector on the
boundary. 
This reflects the absence on the spacetime of a 
globally-defined Killing vector 
that would be timelike in the exterior region 
(cf.~the 
discussion of the isometries of the $\RPthree$ geon in
Sect.~\ref{sec:kruskal-and-geon}). Thus, 
$|\hbox{$\RPtwo$}\rangle$ ``knows'' 
not just about the exterior region of
the $\RPtwo$ geon but also about the region behind the horizons. 

Fourth, the correlations in $|\hbox{$\RPtwo$}\rangle$ 
are between the
right-movers and the left-movers. This is a direct consequence of the
fact that the map 
${\tilde J}_{\rm int}$ on the BTZ hole reverses the spatial
orientation, and 
it reflects thus the spatial nonorientability of the
geon. Fifth, $|\hbox{$\RPtwo$}\rangle$ appears thermal in the
Hawking temperature for operators that do not see the correlations:
this reflects the fact that the geon is a black (and white) hole
spacetime. 

Finally, the expectation value of the energy in
$|\hbox{$\RPtwo$}\rangle$ is, in the limit $a\gg1$, equal to $a^2 / 48
\pi^2$, which is quadratic in $a$ and thus proportional to the ADM
mass of the geon. 
The energy expectation value in the state
$|\hbox{BTZ}\rangle$ on one boundary cylinder of the BTZ hole is also
equal to $a^2 / 48 \pi^2$, for $a\gg1$. In this sense, the energy
expectation value on a single boundary component is the same in
$|0\rangle_{\RPtwo}$ and $|\hbox{BTZ}\rangle$. The analogous property
in the spacetime is that the ADM mass at one infinity is not sensitive
to whether a second infinity exists behind the horizons.

\begin{table}
\caption{Properties of the state $|\hbox{$\RPtwo$}\rangle$, 
and the corresponding properties of 
  the $\RPtwo$ geon spacetime}
\begin{center}
\renewcommand{\arraystretch}{1.4}
\setlength\tabcolsep{5pt}
\begin{tabular}{ll}
\hline\noalign{\smallskip}
$|\hbox{$\RPtwo$}\rangle$ 
  & $\RPtwo$ geon geometry\\
\noalign{\smallskip}
\hline
\noalign{\smallskip}
pure state 
& 
boundary connected 
\\
excited state 
&
not \ads
\\
not static
&
no global KVF 
\\
correlations: left-movers with right-movers
&
spatially nonorientable
\\
right-movers (left-movers) thermal, $T = a/4\pi^2$
&
black hole, $T_H = a/4\pi^2$
\\
$\langle E \rangle = a^2 / 48 \pi^2$, $a\gg1$
&
$M=a^2/(32\pi^2 G_3)$
\\
\noalign{\smallskip}
\hline
\end{tabular}
\end{center}
\label{table:comparison}
\end{table}

\section{Holography and String Theory}%
\label{sec:holography}%
We have seen that the state $|\hbox{$\RPtwo$}\rangle$
on the boundary cylinder of the 
$\RPtwo$ geon mirrors several aspects of the spacetime 
geometry of the $\RPtwo$
geon. Some of this mirroring is immediate from the construction, 
such as
the property that $|\hbox{$\RPtwo$}\rangle$ is a pure state. 
Some aspects of 
the mirroring appear however 
quite nontrivial, especially the fact that the energy
expectation value turned out to be proportional to the ADM mass, and
with the same constant of proportionality as for the vacuum 
$|\hbox{BTZ}\rangle$ on the boundary of the BTZ hole. 
One can see this as a piece of evidence in support of the
holographic hypothesis~\cite{thooft-holo,susskind-holo}, 
according to which physics in the
bulk of a spacetime should be retrievable from physics on the boundary
of the spacetime. 

One would certainly not expect a free conformal scalar field on the
boundary of a spacetime to carry
all the information about the spacetime geometry. 
However, for certain spacetimes related to \ads\ space, a more precise
version of the holographic hypothesis has emerged in string theory 
in the form of the Maldacena duality 
conjectures~\cite{MAST,MAL,malda-more,malda-stillmore}. 
In particular, the 10-dimensional spacetime 
$\CAdSthree\times S^3 \times T^4$, with a flat metric on the 
$T^4$ and a round metric on the~$S^3$, 
is a classical solution to string theory, and the duality conjectures
relate string theory on 
this spacetime to a certain conformal nonlinear sigma-model on
the conformal boundary of the $\CAdSthree$ component. Upon quotienting 
from $\CAdSthree$ to the (in general spinning) BTZ hole, 
the conformal field
theory on the boundary ends in a thermal state analogous to
our $|\hbox{BTZ}\rangle$, but with an energy expectation value that in 
the high temperature limit is not merely 
proportional to but in fact equal to the ADM mass of the
hole~\cite{MAST}. This result can be considered a strong piece of
evidence for the duality conjectures. 

It would now be of obvious interest to adapt our free scalar field
analysis on the boundary of the $\RPtwo$ geon to a string theoretic
context in which the duality conjectures would apply. One would expect
the boundary state again to appear thermal in the Hawking temperature
under some restricted set of observations. The crucial question for
the the holographic hypothesis is how the correlations in the boundary
state might reflect the geometry of the spacetime. As a preliminary
step in this direction, a toy conformal field theory that mimics some
of the anticipated features of the extra dimensions was considered
in~\cite{louko-marolf-btz}, and the energy expectation value in this
toy theory was found to be equal to the geon ADM mass in the high
temperature limit.

\section{Concluding Remarks}%
\label{sec:remarks}%
The results presented here for the $\RPthree$ geon and the $\RPtwo$
geon provide evidence that single-exterior black holes offer a
nontrivial arena for scrutinizing quantum physics
of black holes.  It remains a subject to future work to understand to
what extent the results reflect the 
peculiarities of these particular
spacetimes, and to what extent they might have broader validity. 

In some respects the 
$\RPthree$ and $\RPtwo$ geons are certainly quite nongeneric black
hole spacetimes. 
For example, our quotient constructions on 
Kruskal and the spinless BTZ
hole do not immediately generalize to accommodate spin, 
as the 
putative isometry would need to invert the angular
momentum. 
Similarly, the quotient construction on Kruskal 
does not immediately generalize to the 
Reissner-Nordstr\"om hole, as the relevant isometry would invert 
the electric field. Also,  
the spatial nonorientability of the $\RPtwo$ geon may lead to
difficulties in the string theoretic context. However, 
in 2+1 dimensions there exist locally 
\ads\ single-exterior black (and white) 
hole spacetimes that admit a
spin, and one can choose their spatial topology to be
orientable, for example 
$T^2\setminus${}$\{$point at infinity$\}$
\cite{ABBHP,ABH}. A~natural next step
would be to consider quantum field
theory on these spinning ``wormhole'' spacetimes and on their
conformal boundaries.



\begin{thebibliography}{77}
%
\addcontentsline{toc}{section}{References}

\bibitem{hh-vacuum}
Hartle~J.B., 
Hawking~S.W. 
(1976) 
Phys.\ Rev.\ D {\bf 13}, 2188 

\bibitem{israel-vacuum}
Israel~W. 
(1976) 
Phys.\ Lett.\ {\bf 57A}, 107 

\bibitem{kay-wald}
Kay~B.S., 
Wald~R.M.
(1991) 
Phys.\ Rep.\ {\bf 207}, 49

\bibitem{wald-qft}
Wald~R.M.
(1994)
Quantum Field Theory in Curved Spacetime and Black Hole
Thermodynamics. 
The University of Chicago Press, Chicago

\bibitem{unruh-magnum}
Unruh~W.G. 
(1976) 
Phys.\ Rev.\ D {\bf 14}, 870 

\bibitem{carlip-rev}
Carlip~S. 
(1995)
Class.\ Quantum Grav.\ {\bf 12}, 283 
[gr-qc/9506079]

\bibitem{MAST}  
Maldacena~J., 
Strominger~A. 
(1998)
J.~High Energy Phys.\ {\bf 9812}, 005 
[hep-th/9804085]

\bibitem{horo-marolf-strc}
Horowitz~G.T., 
Marolf~D. 
(1998)
J.~High Energy Phys.\ {\bf 9807}, 014
[gr-qc/9805207]

\bibitem{louko-marolf-btz}
Louko~J., 
Marolf~D. 
(1999)
Phys.\ Rev.\ D {\bf 59}, 066002 
[hep-th/9808081]

\bibitem{Nico} 
Giulini~D. 
(1989)
PhD Thesis, University of Cambridge, Cambridge

\bibitem{topocen}
Friedman~J.L., 
Schleich~K.,
Witt~D.M. 
(1993)
Phys.\ Rev.\ Lett.\ {\bf 71}, 1486; 
Erratum 
(1995)
Phys.\ Rev.\ Lett.\ {\bf 75}, 1872
[gr-qc/9305017]

\bibitem{thooft-holo} 
't~Hooft~G. 
(1993)
In: 
Ali~A., 
Ellis~J.,
Randjbar-Daemi~S. 
(Eds.) 
Salamfestschrift: A~Collection of Talks. 
World Scientific, Singapore
[gr-qc/932102]

\bibitem{susskind-holo}
Susskind~L.
(1995) 
J.~Math.\ Phys.\ {\bf 36} 6377 
[hep-th/9409089]

\bibitem{MAL}  
Maldacena~J. 
(1997)
Adv.\ Theor.\ Math.\ Phys.\ {\bf 2}, 231
[hep-th/9711200]

\bibitem{malda-more} 
Gubser~S.S., 
Klebanov~I.R., 
Polyakov~A.M. 
(1998)
Phys.\ Lett.\ {\bf B428} 105 
[hep-th/9802109]; 
Witten~E. 
(1998)
Adv.\ Theor.\ Math.\ Phys.\ {\bf 2}, 253
[hep-th/9802150];
Claus~P.,
Kallosh~R., 
Kumar~J., 
Townsend~P., 
Van Proeyen~A.
(1998) 
J.~High Energy Phys.\ {\bf 9806}, 004 
[hep-th/9801206];
Susskind~L., 
Witten~E. 
(1998)
e-print hep-th/9805114

\bibitem{malda-stillmore} 
Balasubramanian~V., 
Kraus~P., 
Lawrence~A., 
Trivedi~S.P.
(1998)
e-print hep-th/9808017; 
Keski-Vakkuri~E. 
(1999)
Phys.\ Rev.\ D {\bf 59} 104001
[hep-th/9808037];
Danielsson~U.H., 
Keski-Vakkuri~E., 
Kruczenski~M. 
(1999)
J.~High Energy Phys.\ {\bf 9901}, 002
[hep-th/9812007]; 
Balasubramanian~V., 
Giddings~S.B., 
Lawrence~A.
(1999)
J.~High Energy Phys.\ {\bf 9903}, 001 
[hep-th/9902052]

\bibitem{louko-marolf-geon}
Louko~J.,
Marolf~D. 
(1998)
Phys.\ Rev.\ D {\bf 58}, 024007 
[gr-qc/9802068]

\bibitem{gibb-holo}
Gibbons~G.W.
(1992)
In: 
Kim~J.E. 
(Ed.) 
Proceedings of the 10th Sorak School 
of Theoretical Physics. 
World Scientific, Singapore 

\bibitem{chamb-gibb}
Chamblin~A., 
Gibbons~G.W. 
(1996) 
Phys.\ Rev.\ D {\bf 55}, 2177
[gr-qc/9607079]

\bibitem{GH1}
Gibbons~G.W., 
Hawking~S.W. 
(1977) 
Phys.\ Rev.\ D {\bf 15}, 2752 

\bibitem{BarCarHaw}
Bardeen~J.M., 
Carter~B., 
Hawking~S.W. 
(1973) 
Commun.\ Math.\ Phys.\ {\bf 31}, 161

\bibitem{hawkingCMP}
Hawking~S.W. 
(1975) 
Commun.\ Math.\ Phys.\ {\bf 43}, 199

\bibitem{bekenstein1}
Bekenstein~J.D. 
(1972) 
Nuovo Cimento Lett.\ {\bf 4}, 737

\bibitem{bekenstein2}
Bekenstein~J.D. 
(1974) 
Phys.\ Rev.\ D {\bf 9}, 3292

\bibitem{hawkingCC}
Hawking~S.W. 
(1979)
In: 
Hawking~S.W., 
Israel~W. 
(Eds.) 
General Relativity: An Einstein Centenary Survey. 
Cambridge University Press, Cambridge

\bibitem{ABBHP} 
\AA{}minneborg~S., 
Bengtsson~I., 
Brill~D.R., 
Holst~S.,
Peld\'an~P. 
(1998)
Class.\ Quantum Grav.\ {\bf 15}, 627
[gr-qc/9707036]

\bibitem{BTZ} 
Ba\~nados~M., 
Teitelboim~C., 
Zanelli~J. 
(1992)
Phys.\ Rev.\ Lett.\ {\bf 69}, 1849 
[hep-th/9204099]; 
Ba\~nados~M., 
Henneaux~M., 
Teitelboim~C., 
Zanelli~J. 
(1993)
Phys.\ Rev.\ D {\bf 48}, 1506
[gr-qc/9302012]

\bibitem{birrell-davies}
Birrell~N.D., 
Davies~P.C.W.
(1982)
Quantum Fields in Curved Space. 
Cambridge University Press, Cambridge

\bibitem{takagi-magnum}
Takagi~S.
(1986) 
Prog.\ Theor.\ Phys.\ Suppl.\
{\bf 88}, 1 

\bibitem{ABH} 
\AA{}minneborg~S., 
Bengtsson~I., 
Holst~S. 
(1999) 
Class.\ Quantum Grav.\ {\bf 16}, 363
[gr-qc/9805028]


\end{thebibliography}
\end{document}